# Measuring Bilayer Surface Energy and Curvature in Asymmetric Droplet Interface Bilayers

Nathan E. Barlow,[a,b] Halim Kusumaatmaja,[c] Ali Salehi-Reyhani,[a,b,e] Nick Brooks,[a,b] Laura M. C. Barter,[a,b] Anthony J. Flemming,[d] Oscar Ces[a,b,e,*]

[a] Department of Chemistry, Imperial College London, Exhibition Road, London, SW7 2AZ, UK.
*o.ces@imperial.ac.uk

[b] Institute of Chemical Biology, Imperial College London, Exhibition Road, London, SW7 2AZ, UK

[c] Department of Physics, University of Durham, South Road, DH1 3LE, UK

[d] Syngenta, Jealott's Hill International Research Centre, Bracknell, Berkshire, RG42 6EY, UK

[e] FABRICELL, Imperial College London, London SW7 2AZ

**Abstract:**

For the past decade, droplet interface bilayers (DIBs) have had an increased prevalence in biomolecular and biophysical literature. However, much of the underlying physics of these platforms are poorly characterized. To further the understanding of these structures, a study of lipid membrane tension on DIB membranes is measured by analysing the equilibrium shape of asymmetric DIBs. To this end, the morphology of DIBs is explored for the first time using confocal laser scanning fluorescence microscopy (CLSM). The experimental results confirm that, in accordance with theory, the bilayer interface of a volume asymmetric DIB is curved toward the smaller droplet and a lipid asymmetric DIB is curved toward the droplet with the higher monolayer surface tension. Moreover, the DIB shape can be exploited to measure complex bilayer surface energies. In this study, the bilayer surface energy of DIBs composed of lipid mixtures of phosphatidylgylcerol (PG) and phosphatidylcholine (PC) are shown to increase linearly with PG concentrations up to 25%. The assumption that DIB bilayer area can be geometrically approximated as a spherical cap base is also tested, and it is discovered that the bilayer curvature is negligible for most practical symmetric or asymmetric DIB systems with respect to bilayer area.

## 1. Introduction

Droplet interface bilayers (DIBs)(1) (2) have typically been used to measure membrane bilayer characteristics such as permeability, membrane protein interactions or electrical behaviour. Additionally, interesting DIB morphological behaviour has been studied such as bilayer area modulation by mechanical oscillation,(3) membrane capacitance,(4) or evaporation from the aqueous phase.(5) On a practical level, DIBs have been shown to be particularly useful as they allow for the production of asymmetric membranes,(6) (7) where understanding membrane asymmetry is of high value as it is known to offset transmembrane potential,(8) (9) affect membrane bending rigidity,(10) control membrane protein conformation,(11) as well as membrane permeability.(12) (13) (14)

Surface energy $\gamma$ in bio-membranes is important to quantify as it is known to affect cellular functions such as membrane fusion, ion binding,(15) and integral protein activity.(16) However, measuring surface tension in lipid DIB membranes is challenging, and currently, the only accepted measurement method is made *via* direct visualization of the surface morphology using bright field microscopy, which along with known monolayer surface tensions, can be used to infer bilayer tension. This technique, established by many groups (17) (18) (19) (8) (20) (21) outputs a bilayer surface tension on the order of $\sim 1\ mN\ m^{-1}$ for DIBs made with lipids such as 1,2-diphytanoyl-sn-glycero-3-phosphatidylcholine (DPhPC). For a frame of reference, note that according to Kwok and Evans, the lysis tension for lecithin vesicles was found to be on the order of 3 to 4 $mN\ m^{-1}$.(22) Notably, this high surface tension value (close to known rupture tensions) deviates from that of the vesicular analogue membrane tension, which is often assumed to be negligible.(23) For example, from optical techniques (laser tweezer traps), membrane tethers have been measured to have a surface tension of $3 \times 10^{-3}\ mN\ m^{-1}$.(24) Vesicle fluctuation analysis can also be used to estimate vesicle membrane tension as low as of $10^{-3}\ mN\ m^{-1}$.(25) The surface tension of neutrophils have been calculated to be $0.03\ mN\ m^{-1}$,(26) measured with micropipette aspiration,(27) (28) or the micropipette interfacial area-expansion method.(29) The lipid 1,2-dioleoyl-sn-glycero-3-phospho-(1'-rac-glycerol) (DOPG) was chosen as it is documented that the uncharged PC lipids reduce the surface tension of pulmonary surfactants that contain a large amount of the charged PG lipid.(30) (31)

Certainly, as DIB membranes are high energy systems relative to their vesicular counterparts, measuring membrane tension in DIBs is unfortunately limited by stability. Furthermore, as DIB membrane oscillation cannot be captured optically, and as micropipette aspiration of DIBs would not affect any change in surface tension, it appears that morphological measurements are the only practical option to calculating surface tension. However, though it has been shown that symmetric DIB bilayer surface energies can be estimated using shape information from bright field images, bright field microscopy lacks the ability to capture precise information about membrane curvature due to lipid asymmetry, which can significantly affect the surface energy calculation.

In this study, for the first time using confocal scanning fluorescence microscopy (CSLM) it is shown that, in DIBs composed of droplets of different volumes, there exists curvature in asymmetric bilayers of lipids with differing surface properties. Confocal microscopy was required as it provided a higher resolution image for the bilayer shape which isn't obscured by extraneous light from above and below the midplane of the DIB. This shape information can be applied to the calculation of membrane tension in accordance with a force balance *i.e.* Neumann's triangle(32) (the sine rule). Additionally, a free energy model is applied that describes the curvature behaviour with respect to lipid asymmetry and droplet volume difference.

## 2. Materials and Procedures

### 2.1 Lipid preparation

The lipids 1,2-diphytanoyl-sn-glycero-3-phosphatidylcholine (DPhPC), 1,2-dioleoyl-sn-glycero-3-phospho-(1'-rac-glycerol) (DOPG) and 1-Oleoyl-2-[12-[(7-nitro-2-1,3-benzoxadiazol-4-yl)amino]dodecanoyl]-sn-Glycero-3-Phosphocholine (NDB PC) were purchased from Avanti Polar Lipids. Samples were prepared with 10 mg of solid lipid mixtures suspended in chloroform. The suspension was evaporated where a film was deposited on the vial surface. The film was desiccated for 30 minutes and re-suspended in a $0.25\ M$ phosphate buffer solution at 7.4 pH. The samples were freeze-thaw cycled in liquid nitrogen and in a water bath at 60˚C, repeated 5 times each. The frozen samples were stored at -20˚C until used. Before use, the samples were thawed and diluted to $5\ mg\ mL^{-1}$ and extruded 11 times through $100\ nm$ Avanti PC membrane filters. For confocal microscopy, the fluorescent lipid NBD-PC was similarly deposited on a vial surface where it was suspended in the previously extruded lipid solutions to a molar concentration of 0.1%. It was assumed that the low concentration of NBD-PC does not appreciably affect the surface properties of the lipid monolayer or bilayer.

### 2.2 DIB formation

DIBs were formed by pipetting lipid-in aqueous emulsions into acrylic wells filled with hexadecane. Acrylic wells are chosen for DIB manifolds as the droplet wettability was reduced and has a refractive index of 1.49 (33) which was not dissimilar to the supplier reported value for hexadecane at 1.43. DIBs were formed at 5 $mg\ mL^{-1}$ lipid concentration. The dynamics of monolayer formation have already been established,(34) which show that lipid-in DIBs require a short incubation period on the order of minutes as single droplets in hexadecane before they are pushed together with a needle to form interfaces. There is also a period on the order of minutes where the DIBs 'zip-up' to form a bilayer at equilibrium, curvature measurements are taken at this equilibrium state. Note it is assumed that negligible amounts of residual oil may be trapped in the bilayer, as previous experiments have shown that this DIB system can accommodate the mechanosensitive membrane protein MscL and retain functionality.(35)

### 2.3 Confocal Microscopy

A Leica TCS SP5 confocal fluorescent microscope was used with a 10x objective set with an 84.5 $\mu m$ pinhole (1 airy unit). The field of view was set to 775x775 $\mu m$ (512x512 pixels) and the samples were acquired at a frequency of 400 $Hz$ with 8 line averages. The excitation was achieved with three wavelengths of 458, 476 and 488 $nm$ and absorbance was set at between 510 and 550 $nm$. The images used to fit the model were acquired in the midplane of the droplets. During data collection, focal planes slightly above and below were viewed to confirm that the image was indeed acquired from the midplane.

### 2.4 Pendant Drop Measurements and Drop Shape Analysis (DSA)

It has been shown by Lee *et al* that ionic screening of PC/PG vesicles is required to allow the lipids to coat an air/water monolayer surface.(36) In order to confirm that the lipids have absorbed on the monolayer DSA measurements can also be performed on the lipid solutions in hexadecane. Surface energies of mixtures of lipids were calculated with a pendant tensiometer (Krüss) by drop shape analysis (DSA). The lipids used for making DIBs were formed into aqueous droplets and were immersed in hexadecane from a flat needle 0.52 $mm$ in diameter. The Worthington number $Wo$(37)

$$Wo = \frac{\Delta \rho g V_d}{\pi \gamma D_n} \qquad (2.1)$$

is a dimensionless number which measures the ratio of gravitational to surface forces and is an analogue of the well-known Bond number $Bo = \frac{\Delta \rho g L}{\gamma}$ in bubble systems, where L is the characteristic length.(38) It is often used to estimate the accuracy of the DSA technique, where a measurement is considered accurate at around $Wo \sim 1$ and inaccurate for $Wo \ll 1$.(39) Thus the volume of the droplet must be maximized in order to attain accurate surface energy measurements. For this system, the density difference between water and hexadecane is $\Delta \rho = 230\ kg\ m^{-3}$, acceleration of gravity $g = 9.8\ m\ s^{-2}$, droplet volume $V_d = 0.1 - 0.5 [\times 10^{-9}\ ]m^3$, needle diameter $D_n = 5.2 \times 10^{-4}\ m$ and surface tension $\gamma$ is on the order of $10^{-3}\ J\ m^{-2}$.(39) Due to the low adhesion energy of DPhPC and DOPG monolayers, pendent drop measurements become troublesome as the gravitational potential energy of large droplets overwhelms the pendant droplet adhesion and falls off the needle before equilibrium is reached. This limits the possible range of experimental values of the $Wo$ to between 0.26 to 0.99.

### 3. Results and Discussion

### 3.1 Model Equation and Geometry

It is shown that there may exist a bend in the bilayer between the droplets that form a DIB.(18) Under the assumption that the DIB retains axial symmetry, as demonstrated in **Figure 3.1**, the bilayer bend of radius $r_b$ can be modelled as a section of a spherical cap of height $h_b$ and the droplets themselves can be modelled as intersecting spheres of radius $r_1$ and $r_2$ truncated at height $h_1$ and $h_2$ with spherical cap base radius $a$. Note that there is an important distinction between the effective bilayer curvature $\left(\frac{1}{r_b}\right)$ in a DIB and the lipid spontaneous curvature $c_0$.(40) The curvature in the DIB is a non-local description of the droplet macrostructure. In this work, the lipids DOPG and DPhPC are used as they form stable bilayers(41) with differing surface energies. Though both DPhPC(42) and DOPG(43) have negative spontaneous curvature, planar and positive curvature can occur in DIB membranes.

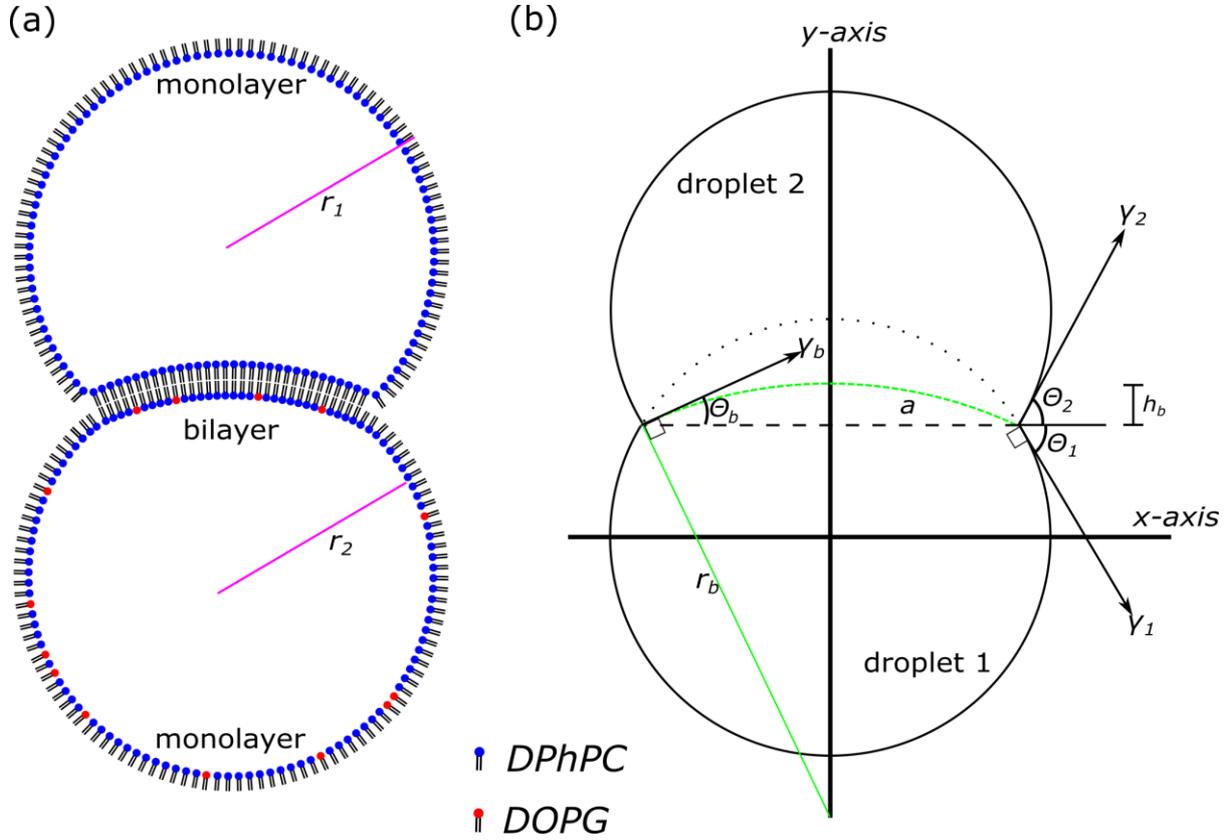

**Figure 3.1**: *Image of (a) cartoon depicting asymmetric DIB with one droplet of radius $r_1$ formed from DPhPC vesicles and one droplet formed from DPhPC doped with DOPG lipids of radius $r_2$. A diagram (b) of an asymmetric DIB which exhibits curvature in the bilayer with surface energy $\gamma_b$ of radius $r_b$ and with angle relative to the x-axis, $\Theta_b$, which balances the surface energies $\gamma_1$ and $\gamma_2$ with contact angles $\Theta_1$ and $\Theta_2$. $h_b$ is the spherical cap height of the bilayer.*

By setting the bilayer concavity toward droplet 2 (see Figure 3.1), due to a surface tension force balance(32) the equations

$$\gamma_1 \cos\Theta_1 + \gamma_2 \cos\Theta_2 = \gamma_b \cos\Theta_b \tag{3.1}$$

$$\gamma_1 \sin\Theta_1 = \gamma_2 \sin\Theta_2 + \gamma_b \sin\Theta_b \tag{3.2}$$

must hold for a given set of bilayer and droplet monolayer surface energies $\gamma_b$, $\gamma_1$ and $\gamma_2$. This is equivalent to the analysis done several authors.(17) (18) (19) (8) (20) (21) The bilayer and droplet contact angles $\Theta_i$ (where the index $i$ is the set $[b, 1, 2]$) are defined in Fig. 3.1(b).

In practice, as droplet radius and position are relatively easy to measure and can be used to measure contact angle $\Theta_1$, $\Theta_2$, and $\Theta$, the force balance of (3.1) and (3.2) can expressed as the equations

$$\gamma_b = \gamma_2 \left( \frac{\sin \theta_2 \cot \theta_1 + \cos \theta_2}{\cos \theta_b - \sin \theta_b \cot \theta_1} \right) \tag{3.3}$$

$$\gamma_b = \gamma_1 \left( \frac{\sin \theta_1 \cot \theta_2 + \cos \theta_1}{\sin \theta_b \cot \theta_2 + \cos \theta_b} \right) \tag{3.4}$$

The usefulness of the form in (3.3-4) becomes apparent if the DIB geometry is known along with single surface energy value $\gamma_1$ or $\gamma_2$, in which case the bilayer surface energy $\gamma_b$ can then be calculated. Thus, based on this a single experimental value of surface energy, both bilayer and monolayer surface energies can be calculated using geometric information from CLSM DIB images. The error propagation analysis of this equation is provided in the ESI.

In section 3.5 we will theoretically consider how DIB asymmetry affect the bilayer area and curvature. To write down a set of equations which are uniquely solvable, we will assume that the volume of the droplets are known and conserved. Using simple geometry, $r_i = \left( \frac{a}{\sin \theta_i} \right)$, and the standard equation for the volume of a spherical cap, the volume of droplets 1 and 2 are

$$V_1 = \frac{\pi}{3} \left( \left( \frac{a}{\sin \theta_1} \right)^3 (2 + 3 \cos \theta_1 - \cos^3 \theta_1) + \left( \frac{a}{\sin \theta_b} \right)^3 (2 - 3 \cos \theta_b + \cos^3 \theta_b) \right) \tag{3.5}$$

$$V_2 = \frac{\pi}{3} \left( \left( \frac{a}{\sin \theta_2} \right)^3 (2 + 3 \cos \theta_2 - \cos^3 \theta_2) - \left( \frac{a}{\sin \theta_b} \right)^3 (2 - 3 \cos \theta_b + \cos^3 \theta_b) \right) \tag{3.6}$$

Now that four equations and four variables remain, namely: equations (3.1), (3.2), (3.5), and (3.6) with variables $\theta_1, \theta_2, \theta_b$, and $a$, the system of equations can be solved. However, as there is no simple analytic solution, this system must be solved using numerical techniques.

### 3.2 Symmetric Lipid DIB Confocal Imaging Result

As an experimental control, symmetric lipid DIBs were formed as shown in **Figure 3.2**. Here the monolayer surface energy of a pure DPhPC monolayer between water and hexadecane is taken as 1.18 $mN\ m^{-1}$.(17) (34). A DIB made up of pure DPhPC with closely matching volumes that vary by less than 1% is shown under CLSM to exhibit no bilayer curvature. To verify that there is no appreciable bilayer curvature, the image is processed with standard techniques using the MATLAB image processing toolbox. All the original data is processed with a Gaussian filter to smooth the edges on the interface peaks and the MATLAB 'fminsearch' function was used to attempt to fit the interface shape to the equation of a circle and to a line. Unsurprisingly, the solver could not fit the interface to the equation of a circle, but could fit to a straight line with a root mean square error (RMSE) of 0.15 depicted as a red line in **Figure 3.2a**. The droplets positions and radii are measured using the MATLAB function "imfindcircles" which employs the Hough(44) transform. The dimensions of the symmetric DIB is **Figure 3.2a** was found to be $r_1 = 433\ \mu m, r_2 = 437\ \mu m, r_b = inf$, and $a = 221\ \mu m$. From equations (3.3) and (3.4), with the input value of $\gamma_1 = \gamma_2 = 1.18\ mN\ m^{-1}$, the bilayer surface energy was calculated to be $\gamma_b = 2.04\ mN\ m^{-1}$ with an error of $0.121\ mN\ m^{-1}$ (see ESI for error propagation analysis), matching previously reported surface energy results from Taylor.(17)

In contrast, a non-similar volume DIB (*i.e.* a volume ratio of 0.37) is shown to exhibit a circular curve in the bilayer which bends toward the smaller droplet, shown in **Figure 3.2b**. To calculate the bilayer curvature, the image is processed again on MATLAB using the image processing. Within the region of interest (ROI), the maximum intensity peak values are obtained along the vertical axis. These peak values are fit to the equation of a circle using the MATLAB function "fminsearch" to minimize the root mean squared error (RMSE) of the distance from a peak point to the fit circle. The droplet dimensions are also measured using the MATLAB function "imfindcircles". From this the ratio of the bilayer radius of curvature to the smaller droplet radius of curvature in the figure is measured to be 7.21 with a RMSE of 0.12 depicted as a red line. Based on the measured, normalized geometry of $r_1 = 397\ \mu m, r_2 = 535\ \mu m, r_b = 2859\ \mu m$, and $a = 270\ \mu m$, the surface energy for the bilayer is calculated to be $\gamma_b = 1.93\ mN\ m^{-1}$ with an error of $0.107\ mN\ m^{-1}$. Here actual bilayer surface energy measurement is within error of the previously reported measurement, $2.04\ mN\ m^{-1}$.(17)

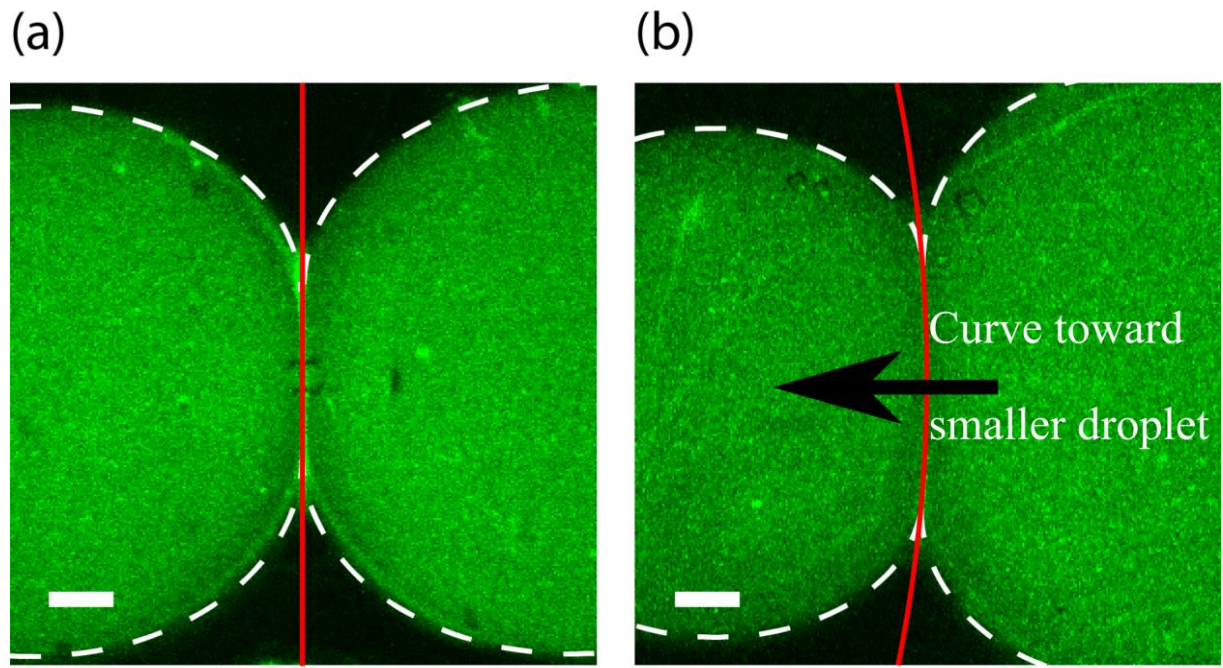

**Figure 3.2**: *Filtered CLSM image (a) of a volume symmetric DIB formed from a single lipid type. Filtered image (b) of a volume asymmetric DIB (volume ratio of 0.37) that exhibits bilayer interface curvature. Both droplets consist of DPhPC with dilute (0.1% molar fraction) NDB-PC dye formed in hexadecane. The bilayer concavity faces the smaller droplet with higher Laplace pressure. Scale bars are 100 μm.*

### 3.3 Asymmetric Lipid DIB Confocal Imaging Result

The results of the CLSM experiment on asymmetric DIBs is provided in **Figure 3.3**. Three DIBs of varying degrees of monolayer asymmetry, from lowest to highest, are shown to exhibit bilayer curvature, where the interface curvature is measured with a MATLAB script where the derivative of the fluorescence intensity plot is used to find the image edge threshold which is fit to the equation of a circle by minimizing the RMSE, see ESI for more details. The geometry furthermore can be used to calculate the bilayer and the monolayer surface energy. Note that in the following cases the pure DPhPC lipid droplet surface energy is assumed to remain $\gamma_2 = 1.18\ mN\ m^{-1}$. **Figure 3.3(a)** shows a bilayer curvature to droplet curvature ratio of 4.96 and a spherical cap base radius to droplet radius ratio of 0.44 at a RMSE of 0.31. Note that in **Figure 3.3(a)** the monolayer in the dark (leftmost) droplet 1 is composed of 6% DOPG from total lipid content, which is left dark to enhance the contrast in the bilayer threshold. The geometric measurements of the DIB are $r_1 = 431\ \mu m, r_2 = 431\ \mu m, r_b = 2138\ \mu m$, and $a =$

194 $\mu m$, where a brightfield image of the dark droplet is used to measure the dimensions of the dark droplet. The increased surface energy for droplet 1 and the bilayer is calculated to be $\gamma_1 = 1.70\ mN\ m^{-1}$, and $\gamma_b = 2.58\ mN\ m^{-1}$ with an error of $0.149\ mN\ m^{-1}$.

Further increasing the DIB asymmetry shown in **Figure 3.3(b)** confirms that the bilayer radius of curvature ratio deceases to 3.34 with a spherical cap base radius to droplet radius ratio of 0.49 at a RMSE of 0.33. The asymmetric DIB is composed of 12% DOPG from total lipid content in the left droplet with dimensions measured to be slightly volume asymmetric, $r_1 = 452\ \mu m$, $r_2 = 428\ \mu m$, $r_b = 1420\ \mu m$, and $a = 225\ \mu m$. Similarly, the surface energy for droplet 1 and the bilayer is calculated to be $\gamma_1 = 1.92\ mN\ m^{-1}$, and $\gamma_b = 2.68\ mN\ m^{-1}$ with an error of $0.169\ mN\ m^{-1}$.

The third and highest stable asymmetric DIB formed in **Figure 3.3(c)** is composed of 25% DOPG from total lipid content. The bilayer radius of curvature ratio is measured at 2.23 and spherical cap base radius to droplet radius ratio of 0.58 with a RMSE of 0.516. The DIB dimensions are calculated to be $r_1 = 334\ \mu m$, $r_2 = 331\ \mu m$, $r_b = 738\ \mu m$, and $a = 185\ \mu m$, where the surface energy for droplet 1 and the bilayer is calculated to be $\gamma_1 = 2.70\ mN\ m^{-1}$, and $\gamma_b = 3.33\ mN\ m^{-1}$ with an error of $0.195\ mN\ m^{-1}$.

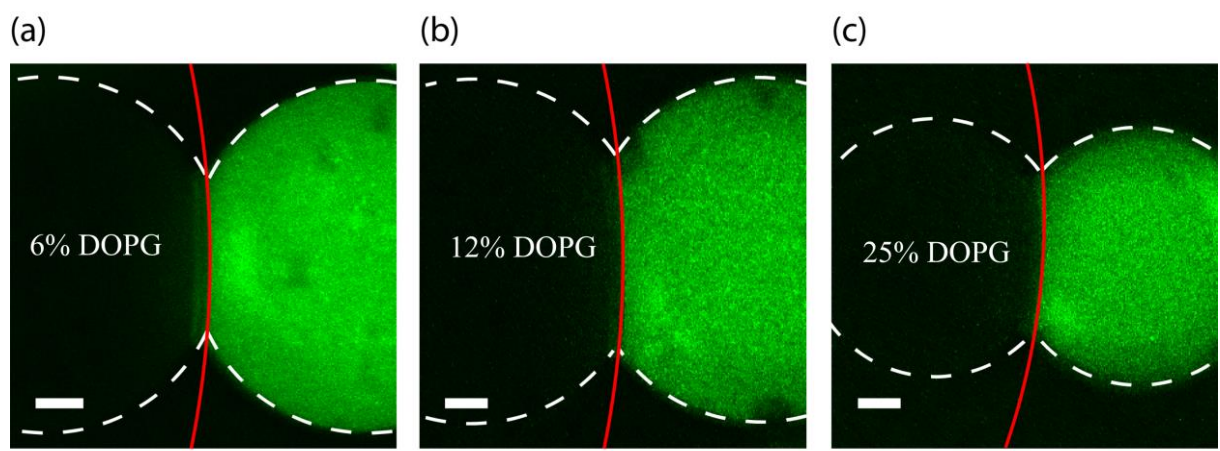

**Figure 3.3**: *The filtered CLSM images of matching volume droplet DIBs in hexadecane with mismatched surface energies exhibits curvature in the bilayer towards the higher surface tension droplet. The dark droplet contains the DOPG and light droplet contains the DPhPC with NDBPC. The amount of DOPG in images (a), (b) and (c) are respectively 6%, 12% and 25% from total lipid content. Scale bars are 100 μm.*

### 3.4 Pendant Drop Measurements (DSA) Results

To compare with the above DIB method of measuring surface energy, the results from the DSA measurements are provided in **Figure 3.4**. The measurement was taken for droplets that could attain equilibrium without falling from the flat syringe needle. Note that the Worthington number ($Wo$) is close to 1 for most measurements. The results below can be used to verify the DIB morphology method for surface tension.

| Percent DOPG | $\gamma_1, mN\ m^{-1}$ | STDEV | $Wo$ |
|---|---|---|---|
| 50.0 | 3.13 | 1.56 | 0.26 |
| 25.0 | 2.65 | 1.35 | 0.71 |
| 12.5 | 2.02 | 1.02 | 0.67 |
| 6.3 | 1.74 | 0.91 | 0.99 |
| 0.0 | 1.50* | 0.82 | 0.70 |

**Figure 3.4**: *Table of surface energy measurements for a droplet of DPhPC with a given DOPG% that forms a monolayer between water and hexadecane. The error and Worthington number are provided for reference. *Note that the literature value for pure DPhPC is 1.18 $mN\ m^{-1}$.(17) (34).*

The DSA results show good agreement with the DIB method as shown in **Figure 3.5**. This verifies the technique developed by several authors with respect to symmetric DIBs.(17) (18) (19) (8) (20) (21) However, here we have shown that with the use of CLSM, we can capture bilayer curvature data to be used in calculating asymmetric bilayer surface energy. This is useful as with significantly low surface energies (lower than 5 $mN\ m^{-1}$) it is often difficult to obtain shape measurements from the standard DSA method as the droplets tend to fall from the needle.(34) Given that they are stable and stationary, by the DIB method, asymmetric bilayer surface energies can be calculated. We further note that the discrepancy in the measurement of pure DPhPC comes chiefly from the fact that DSA measurements become more difficult with low surface tensions, in this case $\gamma$ lower than 1.5 $mN\ m^{-1}$.

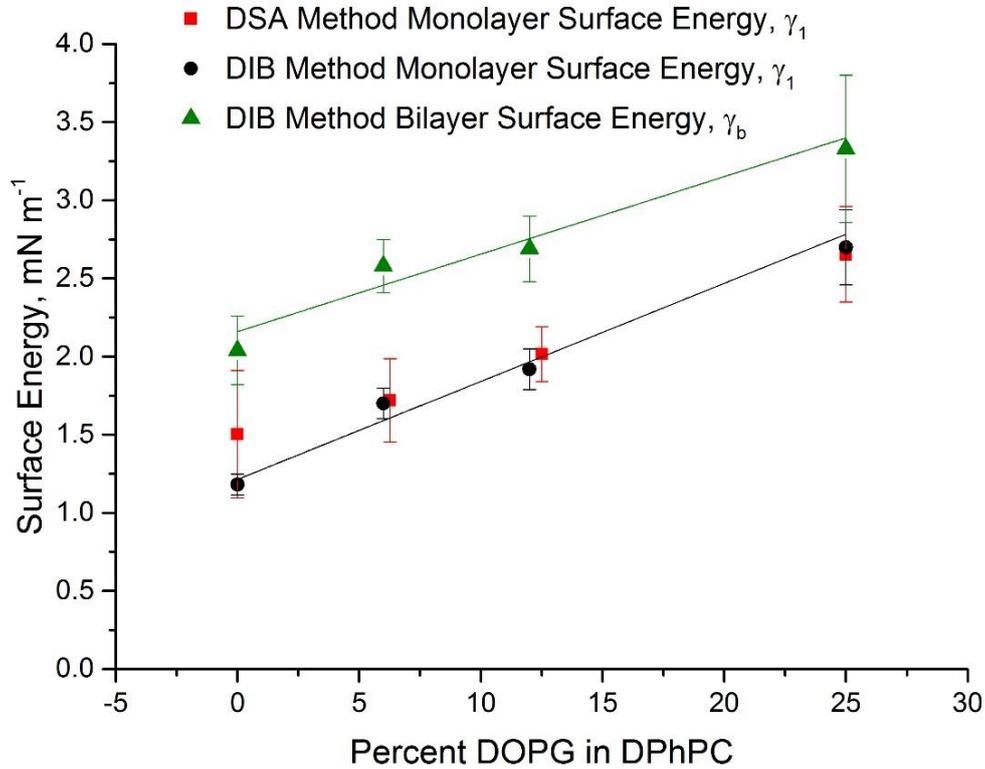

**Figure 3.5**: *Bilayer and monolayer surface energies obtained from DSA and DIB methods as a function of monolayer asymmetry in DIBs with a droplet composed of pure DPhPC and a droplet with a mixture of DOPG in DPhPC. For the DIB method, the droplet 2 surface energy is assumed to be $\gamma_2 = 1.18\ mN\ m^{-1}$ for pure DPhPC. The linear fits for the bilayer and monolayer surface energies have a Pearson's R-squared value of 0.94 and 0.98 respectively with a sample size is $n = 3$.*

**3.5 Droplet Morphology Model Result**

The free energy model described in **Section 3.1** was applied to investigate asymmetric and symmetric DIB morphology with the given system surface energies $\gamma_1$, $\gamma_2$, and $\gamma_b$ by equation (3.1), (3.2), (3.5) and (3.6). A simple way to analyse the system is to view the interface diameter $a$ normalized by the droplet radius $r_m$. This is useful as it can be generalized and scaled for different droplet systems driven by surface energy minimization. By this assessment, the symmetric model is $\frac{a}{r_m} = \sqrt{\left(1 - \left(\frac{\gamma_b}{\gamma_m}\frac{1}{2}\right)^2\right)}$ which is shown by **Figure 3.6**, where the DIB monolayer surface energies $\gamma_1$ and $\gamma_2$ of droplets 1 and 2 are the reference values, i.e.: $\gamma_b$ is in the form of $\frac{\gamma_b}{\gamma_m}$. As the bilayer surface energy is decreased from $\frac{\gamma_b}{\gamma_m} = 2$ or $\left(\frac{a}{r_m} = 0\right)$ the DIB will start to 'zip up'. This 'zip up' process is defined as an increase in contact surface area between droplets. Here one can see that the ratio of spherical cap base radius $a$ to droplet radius $r_1$ and $r_2$ increase drastically following the arrow and gradually decreases until $\frac{\gamma_b}{\gamma_m} = 0$ or $\left(\frac{a}{r_m} = 1\right)$. This is an unsurprising result as it has already been shown by (3.1) that the contact angle $\Theta_m$ changes as $\cos\Theta_m = \frac{\gamma_b}{2\gamma_m}$. This 'zip up' process occurs mainly up to the point where the bilayer surface energy matches that of the monolayers, or $\gamma_m = \gamma_b$ or $\left(\frac{\gamma_m}{\gamma_b} = 1\right)$. After this point any small perturbation in bilayer surface energy will have a diminished effect on bilayer radius $a$. Note that as surface energy is finite, the DIB can only 'zip up' completely if $\gamma_b = 0$.

There is no simple analytical solution to the asymmetric case. However, it can be solved using numerical techniques. A MATLAB script was employed to solve for the variables $r_1, r_2, r_b$ and $a$. The script employs the 'fmincon' function, which runs an 'interior-point' algorithm, to solve for the minimization of the free energy functional(21) $f$ of surface energy $\gamma$ and surface area $A$,

$$f = \gamma_1 dA_1 + \gamma_2 dA_2 + \gamma_b dA_b \tag{3.7}$$

under the constraint that $V_1$ and $V_2$ (from equation 3.7 and 3.8) are constant. This script was used to solve for the ratio of the drop radii $r_1, r_2$ and the bilayer radius $r_b$. **Figure 3.7** shows that, for asymmetric DIBs, the membrane radius $r_b$ will decrease with increasing asymmetry in monolayer surface tension $\frac{\gamma_1}{\gamma_2}$ until it matches the spherical cap base radius, or $r_b = a$. For simplicity, here we have used $\gamma_2 = \gamma_b$ as the reference tension value. Note that a DIB with a bilayer of infinite radius $r_b \to \infty$ (zero mean curvature) exists at the symmetric limit. The model result also shows that for asymmetric DIBs greater than the range of $\frac{\gamma_1}{\gamma_2} \sim 1.2$, small changes in asymmetry affect the bilayer radius of curvature significantly.

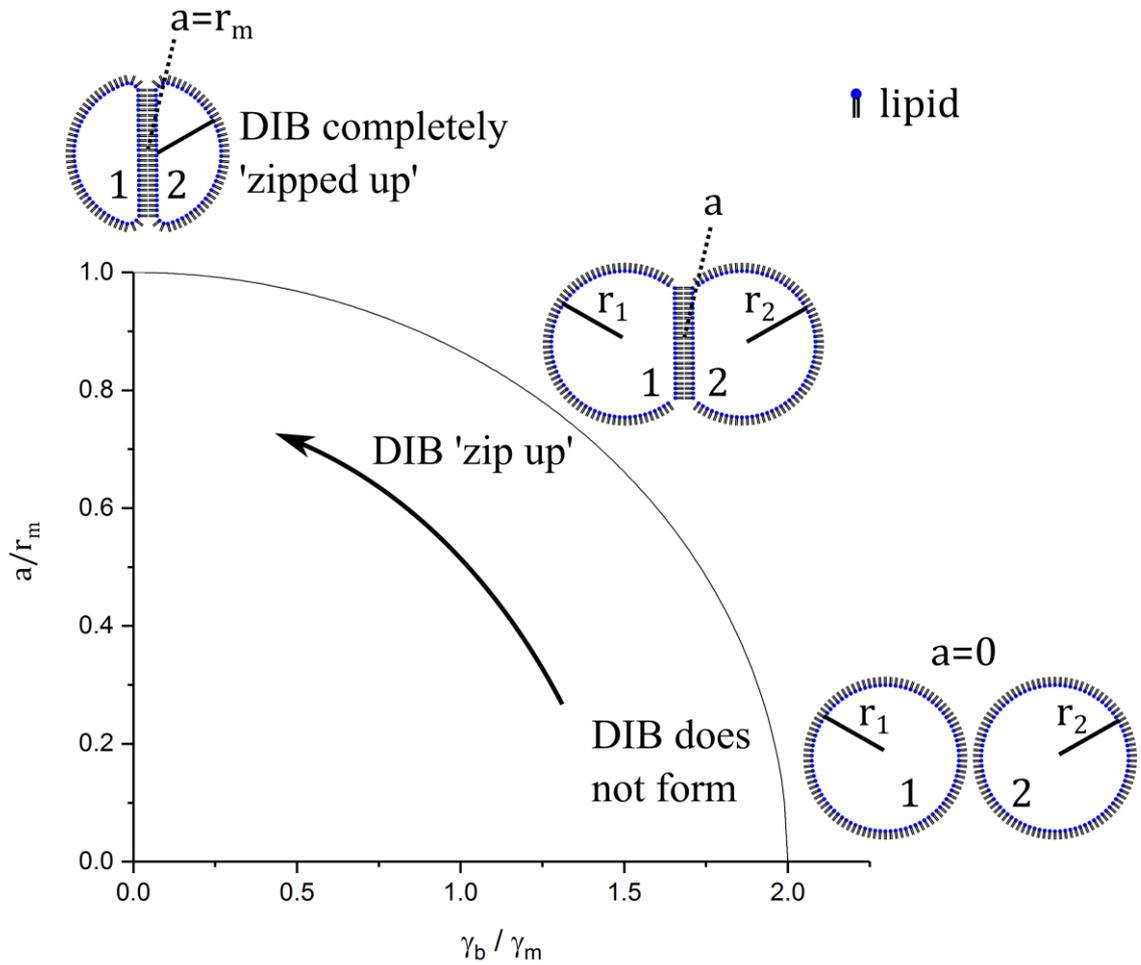

**Figure 3.6**: *Symmetric DIB model result for the ratio between the spherical cap base radius $a$ and the droplet radii $r_1$ and $r_2$ as a function of bilayer to monolayer tension $\gamma_b/\gamma_m$. The droplets 'zip up' drastically with increasing monolayer surface energy up to the point the bilayer and monolayer energies match, after which the effect is less dramatic until the droplets 'zip up' completely and the droplet radius matches that of the bilayer radius. Note that the lipids in the DIB cartoons are for shape reference and not drawn to scale.*

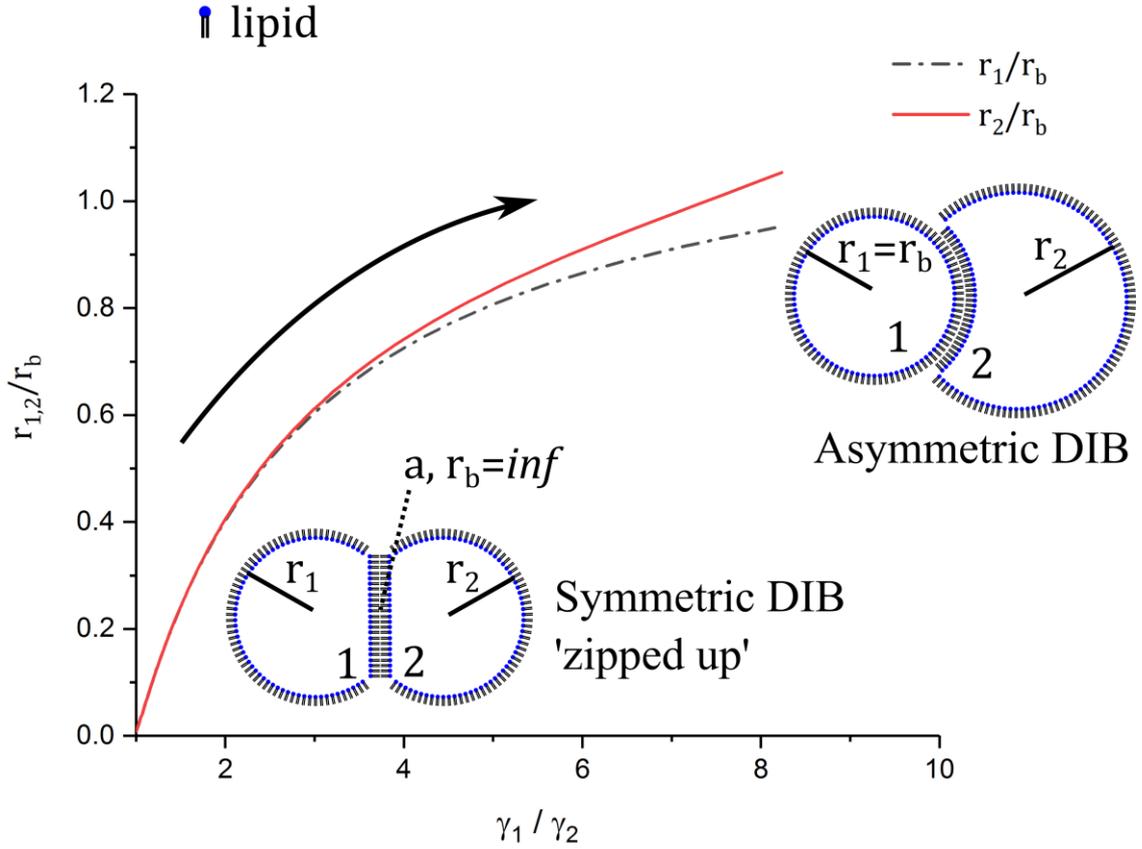

**Figure 3.7**: *Asymmetric DIB model result for the ratio of the droplet radii $r_1$ and $r_2$ to bilayer radius $r_b$ as a function of monolayer tension ratio $\gamma_1/\gamma_2$. Here $\gamma_2 = \gamma_b$. Due to the force balance, the bilayer deviates from the spherical cap base radius $a$, and following the arrow, it will continue to curve towards the higher tension droplet until the bilayer radius matches that of the droplet radius $r_1 = r_b$. Due to a mass balance, the droplet diameter of 2 will increase above that of droplet 1. Note that the lipids are not drawn to scale in the DIB cartoon.*

Often bilayer area is approximated by the spherical cap base radius (or the linear distance between the intersecting circles) $A_{approx} = \pi a^2$. The definitions of bilayer area considering curvature (spherical cap area) is given as $A_b = 2\pi r_b h_b$. Therefore, the percent area deviation from the linear approximation can be defined as $\Delta A$,

$$\Delta A = \left(\frac{2\pi r_b h_b}{\pi a^2} - 1\right) \times 100\% \qquad (3.8)$$

As shown in **Figure 3.8** for volume symmetric DIBs, increasing the monolayer asymmetry elicits only a modest deviation in surface area. However, if the droplet volume asymmetry is modified the area deviation can be magnified. Note that typically DIB droplets are roughly the same size, and high surface energy asymmetry does not appear to be stable experimentally. By applying this model, the area correction of the DIB bilayers found experimentally via CLSM can be determined. For the volume asymmetric droplet (**Figure 3.2b**), by equation 3.10, the area deviation $\Delta A$ is found to be 0.22%. Additionally, the area deviation for the lipid asymmetric DIBs is found to be 0.21, 0.63, and 1.6% for the 6, 12 and 25% DOPG in DPhPC respectively. This shows that, at least for the range of DIB asymmetry explored in this study, the linear approximation of area is a reasonable estimate. Indeed, according to **Figure 3.8**, even relatively high monolayer asymmetry manifests as a deviation of less than 5% for volume symmetric DIBs.

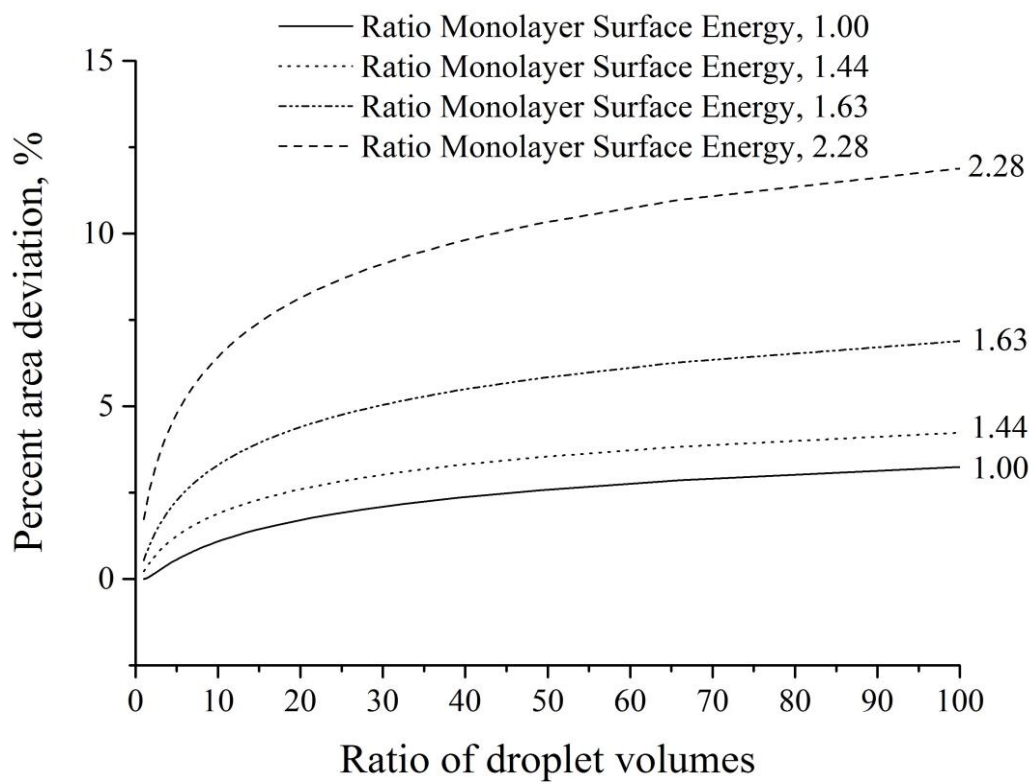

**Figure 3.8**: *Model result of the percent area deviation of DIBs with volume and lipid asymmetry given that the bilayer surface tension is set to the average of the two monolayer surface tensions. The effect of monolayer surface energy is exacerbated by increasing droplet volume differences.*

### 3.6 Model and System Limitations

### 3.6.1 Practical Limitations of the Method

The valid range of $\gamma_1$ and $\gamma_2$ for the model is limited by DIB stability as experimentally stable DIBs are only formed below a surface energy ratio of 2.5. Above this level the droplets are disposed to coalesce into one larger droplet. This can be explained by the fact that emulsion or DIB stability depends on a) the osmotic and Laplace pressure of the droplets and b) on the pressure balance across the membrane.(45) Inescapably, the difference in pressure between connecting droplets may lead to inherent instability for large droplet volume ratios which limits the practicality for extreme droplet volume ratios.

### 3.6.2 Limitations in Scalability

The use of DIBs for measuring surface tension is limited in size to micro scale droplets. This is the case as other thermodynamic factors come into play on smaller length scales such as line tension, which becomes non-negligible once the droplet reaches length scales below $100\ nm$.(46) Additionally, it is important to note that attaining thermodynamic equilibrium can be somewhat troublesome for DIBs as they continually lose water mass due to evaporation, this is a particular problem for DIBs with diameters on a micron length scale.(5) The effect of evaporation on DIBs was characterized recently by Venkatesan *et al*,(47) and this effect is mitigated by using droplets on the order of 300 micron in diameter covered by a thick layer of oil, however the effect of gravity on droplet shape prevents the use of much larger DIBs without adding another level of complexity to the model.(48) Indeed, the model system is limited in scalability by the Bond number (see Section 2.4). For this study the density difference between water and hexadecane is $\Delta\rho = 230 \ kg \ m^{-3}$, acceleration of gravity $g = 9.8 \ m \ s^{-2}$, and surface tension $\gamma$ is on the order of $10^{-3} \ J \ m^{-2}$. If the characteristic length is taken as droplet radius, then $L$ is on the order of $2.3 - 4.0 \ [\times 10^{-4}]m$, this implies a Bond number of approximately 0.1 to 0.4. A reasonable upper limit for DIB applications is a Bond number less than 1 (a droplet radius of 660 micrometres for the system at hand), as values greater than 1 imply a decreased effect of surface tension relative to gravity and will result in non-spherical droplets. Note that this model does not account for non-spherical droplets.

### 3.6.3 Model limitations

The model is limited to static equilibria and cannot be used to probe the absolute surface tension of the DIB membrane out of equilibrium, though the relative surface forces, such as $\frac{\gamma_b}{\gamma_2} = f(\Theta_1, \Theta_2, \Theta_b)$, can be calculated from equations (3.3) and (3.4). In a similar vein, the model is only valid for systems that are under tension. More specifically, this model would not be particularly useful to measure the tension of adhering vesicles, as the mechanical tension is not necessarily known as the bodies can be deflated and the energetics can be affected by the expansion modulus.(49)

## 4. Conclusions

For the first time it has been shown that asymmetric DIBs form curved surface in the bilayer due to a surface energy balance. It is analogous to the effect of volume differences, but here the surface energy asymmetry controls this behaviour. As shown by Taylor *et al* for symmetric DIBs(17), our study shows that the curvature effect in asymmetric DIBs can be employed as an alternative method of measuring interfacial tension of complex, asymmetric lipid monolayers or bilayers through the application of CLSM. The obtained interfacial tension values are in good agreement with droplet shape analysis results. Furthermore, the results obviate the negligible effect of area deviation with respect to DIB asymmetry; though the effect of curvature strongly affects the surface tension calculation, even the most asymmetric system in this experiment (with a surface energy ratio of approximately 2.2) corresponds to a deviation of only 1.6%. Thus, with DIB platforms, the bilayer interfacial area measurement is only affected by the extreme cases of high surface tension asymmetry and extreme volume mismatch, an important validation of an assumption made in many published DIB applications.

A linear relationship between bilayer surface energy with respect to DOPG and DPhPC mixtures is shown up to 25% DOPG. However, this linear relationship is not necessarily the case for all lipid mixtures. For example, significant non-linearity and hysteresis in dynamic interfacial tension measurements as a function of the mole fraction of cholesterol in lecithin lipids has been observed.(50) The formation of a lipid-lipid complex has been shown for phosphatidylcholine-phosphatidylethanolamine and sphingomyelin-ceramide mixtures, this implies a non-linear relationship for interfacial tension with respect to lipid concentrations.(51) Thus, the asymmetric DIB morphology method could be used to probe this non-linear surface tension behaviour by measuring the surface morphology as a function of lipid content and asymmetry.

There is a wide range of possibilities for future work measuring surface tension and curvature affects in DIBs, giant unilameller vesicles (GUVs)(52) or even cells. It has already been shown that curvature exists between adhering cells as observed in the biologically mediated cell-cell contact between *C. elegans* embryos(53) and between adhering vesicles.(42) Investigations of lipid flip-flop in bio-membranes,(54) to a marginal degree of success, have been performed using sum frequency vibrational spectroscopy,(55) indirectly with ceramide induced trans-bilayer movement in vesicles,(54) small angle neutron scattering,(56) and by molecular simulation.(57) Asymmetric DIBs or adhering vesicles offer an alternative measurement technique for the rate of lipid flip-flop, by directly measuring the decrease of interfacial curvature as the lipids flip from one droplet or vesicle to another. Here the challenge lies in distinguishing the rate of flip-flop from the rate of lateral lipid diffusion(58) between the monolayer and bilayer, as well as lipid uptake into the bilayer.(5, 50, 59) However, a recent publication has demonstrated a promising technique for determining bilayer flip-flop on DIB membranes via parallel capacitance based measurements on an integrated microfluidic device; in this study it was successfully shown that surface bound peptides (alamethicin) facilitate the movement of lipids between leaflets.(9)

The present model is valid for stationary surfaces at equilibrium. It would also be interesting to extend the model to dynamic behaviour of micro-DIBs where the droplets change shape and the bilayer may even buckle.(60) The bilayer buckling indicates that the effective bilayer surface tension $\gamma_b$ had dropped to zero.(5) Understanding interfacial physical chemistry is paramount to the development of DIBs as a tool for biological discovery, which is crucial for burgeoning fields such as synthetic biology and biotechnology.

**Competing Interests**

The authors declare no competing financial interests.

**Authors' Contributions**

N.E.B. performed experiment and model analysis, wrote the main text of the document. H.K. edited the text, discussed the model and data analysis, A.S-R. edited text, built the experimental setup and discussed the results, O.C., A.J.F, L.M.C.B and N.B. edited text and discussed results.

**Funding**

The research leading to these results has received funding from the European Union Seventh Framework Programme (FP7/2007-2013) under grant agreement nº 607466. This work was supported by an Imperial College Fellowship awarded to A.S-R. This research was funded by and BBSRC EPSRC grants: EP/J017566/1, EP/L015498/1, EP/J021199/1 and EP/K503733/1.

**References**

1. Hagan Bayley BC, Andrew Heron, Matthew A. Holden, William Hwang, Ruhma, Syeda JT, and Mark Wallace. Droplet Interface Bilayers. Molecular Biosystems. 2008;4(12):1191-208.
2. Funakoshi K, Suzuki H, Takeuchi S. Lipid bilayer formation by contacting monolayers in a microfluidic device for membrane protein analysis. Analytical chemistry. 2006;78(24):8169-74.
3. Najem JS, Freeman EC, Yasmann A, Sukharev S, Leo DJ. Mechanics of Droplet Interface Bilayer "Unzipping" Defines the Bandwidth for the Mechanotransduction Response of Reconstituted MscL. Advanced Materials Interfaces. 2016.
4. Gross LC, Heron AJ, Baca SC, Wallace MI. Determining membrane capacitance by dynamic control of droplet interface bilayer area. Langmuir. 2011;27(23):14335-42.
5. Mruetusatorn P, Boreyko JB, Venkatesan GA, Sarles SA, Hayes DG, Collier CP. Dynamic morphologies of microscale droplet interface bilayers. Soft Matter. 2014;10(15):2530-8.

6. Hwang WL, Chen M, Cronin B, Holden MA, Bayley H. Asymmetric Droplet Interface Bilayers. Journal of the American Chemical Society. 2008;130(18):5878-9.
7. Nguyen M-A, Sarles SA, editors. Microfluidic Generation, Encapsulation and Characterization of Asymmetric Droplet Interface Bilayers. ASME 2016 Conference on Smart Materials, Adaptive Structures and Intelligent Systems; 2016: American Society of Mechanical Engineers.
8. Freeman EC, Najem JS, Sukharev S, Philen MK, Leo DJ. The mechanoelectrical response of droplet interface bilayer membranes. Soft Matter. 2016;12(12):3021-31.
9. Taylor G, Nguyen M-A, Koner S, Freeman E, Collier CP, Sarles SA. Electrophysiological interrogation of asymmetric droplet interface bilayers reveals surface-bound alamethicin induces lipid flip-flop. Biochimica et Biophysica Acta (BBA)-Biomembranes. 2018.
10. Elani Y, Purushothaman S, Booth PJ, Seddon JM, Brooks NJ, Law RV, et al. Measurements of the effect of membrane asymmetry on the mechanical properties of lipid bilayers. Chemical Communications. 2015;51(32):6976-9.
11. Traïkia M, Warschawski DE, Lambert O, Rigaud J-L, Devaux PF. Asymmetrical Membranes and Surface Tension. Biophysical Journal. 2002;83(3):1443-54.
12. Hill WG, Rivers RL, Zeidel ML. Role of Leaflet Asymmetry in the Permeability of Model Biological Membranes to Protons, Solutes, and Gases. The Journal of General Physiology. 1999;114(3):405-14.
13. Krylov AV, Pohl P, Zeidel ML, Hill WG. Water permeability of asymmetric planar lipid bilayers: leaflets of different composition offer independent and additive resistances to permeation. J Gen Physiol. 2001;118(4):333-40.
14. Milianta PJ, Muzzio M, Denver J, Cawley G, Lee S. Water Permeability across Symmetric and Asymmetric Droplet Interface Bilayers: Interaction of Cholesterol Sulfate with DPhPC. Langmuir. 2015;31(44):12187-96.
15. Ohki S, Ohshima H. Divalent cation-induced phosphatidic acid membrane fusion. Effect of ion binding and membrane surface tension. Biochimica et Biophysica Acta (BBA)-Biomembranes. 1985;812(1):147-54.
16. Lee AG. How lipids affect the activities of integral membrane proteins. Biochimica et Biophysica Acta (BBA)-Biomembranes. 2004;1666(1):62-87.
17. Taylor GJ, Venkatesan GA, Collier CP, Sarles SA. Direct in situ measurement of specific capacitance, monolayer tension, and bilayer tension in a droplet interface bilayer. Soft Matter. 2015;11(38):7592-605.
18. Dixit SS, Pincus A, Guo B, Faris GW. Droplet Shape Analysis and Permeability Studies in Droplet Lipid Bilayers. Langmuir : the ACS journal of surfaces and colloids. 2012;28(19):7442-51.
19. Yanagisawa M, Yoshida T-a, Furuta M, Nakata S, Tokita M. Adhesive force between paired microdroplets coated with lipid monolayers. Soft Matter. 2013;9(25):5891-7.
20. Kancharala AK, Freeman E, Philen MK, editors. Energy harvesting from droplet interface bilayers. ASME 2015 Conference on Smart Materials, Adaptive Structures and Intelligent Systems; 2015: American Society of Mechanical Engineers.
21. Villar G, Heron AJ, Bayley H. Formation of droplet networks that function in aqueous environments. Nat Nano. 2011;6(12):803-8.
22. Kwok R, Evans E. Thermoelasticity of large lecithin bilayer vesicles. Biophysical journal. 1981;35(3):637-52.
23. Jähnig F. What is the surface tension of a lipid bilayer membrane? Biophysical Journal. 1996;71(3):1348-9.
24. Hochmuth F, Shao J-Y, Dai J, Sheetz MP. Deformation and flow of membrane into tethers extracted from neuronal growth cones. Biophysical journal. 1996;70(1):358-69.
25. Rädler JO, Feder TJ, Strey HH, Sackmann E. Fluctuation analysis of tension-controlled undulation forces between giant vesicles and solid substrates. Physical Review E. 1995;51(5):4526.
26. Hochmuth RM. Micropipette aspiration of living cells. Journal of biomechanics. 2000;33(1):15-22.

27. Guevorkian K, Colbert M-J, Durth M, Dufour S, Brochard-Wyart F. Aspiration of biological viscoelastic drops. Physical review letters. 2010;104(21):218101.
28. Evans E, Rawicz W, Smith BA. Back to the future: mechanics and thermodynamics of lipid biomembranes. Faraday Discussions. 2013;161:591-611.
29. Kinoshita K, Parra E, Needham D. New Sensitive Micro-Measurements of Dynamic Surface Tension and Diffusion Coefficients: Validated and Tested for the Adsorption of 1-Octanol at a Microscopic Air-Water Interface and its Dissolution into Water. Journal of Colloid and Interface Science. 2017.
30. Klenz U, Saleem M, Meyer M, Galla H-J. Influence of lipid saturation grade and headgroup charge: a refined lung surfactant adsorption model. Biophysical journal. 2008;95(2):699-709.
31. Schürch S, Goerke J, Clements JA. Direct determination of surface tension in the lung. Proceedings of the National Academy of Sciences. 1976;73(12):4698-702.
32. Rowlinson JS, Widom B. Molecular Theory of Capillarity: Dover Publications; 1982.
33. Kasarova SN, Sultanova NG, Ivanov CD, Nikolov ID. Analysis of the dispersion of optical plastic materials. Optical Materials. 2007;29(11):1481-90.
34. Venkatesan GA, Lee J, Barati Farimani A, Heiranian M, Collier CP, Aluru NR, et al. Adsorption kinetics dictate monolayer self-assembly for both lipid-in and lipid-out approaches to droplet interface bilayer formation. Langmuir. 2015.
35. Barriga HMG, Booth P, Haylock S, Bazin R, Templer RH, Ces O. Droplet interface bilayer reconstitution and activity measurement of the mechanosensitive channel of large conductance from Escherichia coli. Journal of the Royal Society Interface. 2014;11(98).
36. Lee S, Kim DH, Needham D. Equilibrium and Dynamic Interfacial Tension Measurements at Microscopic Interfaces Using a Micropipet Technique. 2. Dynamics of Phospholipid Monolayer Formation and Equilibrium Tensions at the Water–Air Interface. Langmuir. 2001;17(18):5544-50.
37. Worthington AM. On pendent drops. Proceedings of the Royal Society of London. 1881;32(212-215):362-77.
38. Yang L, Kapur N, Wang Y, Fiesser F, Bierbrauer F, Wilson MC, et al. Drop-on-demand satellite-free drop formation for precision fluid delivery. Chemical Engineering Science. 2018.
39. Berry JD, Neeson MJ, Dagastine RR, Chan DYC, Tabor RF. Measurement of surface and interfacial tension using pendant drop tensiometry. Journal of Colloid and Interface Science. 2015;454:226-37.
40. McMahon HT, Boucrot E. Membrane curvature at a glance. Journal of Cell Science. 2015;128(6):1065-70.
41. Yasmann A, Sukharev S. Properties of Diphytanoyl Phospholipids at the Air–Water Interface. Langmuir. 2015;31(1):350-7.
42. Sakuma Y, Imai M, Yanagisawa M, Komura S. Adhesion of binary giant vesicles containing negative spontaneous curvature lipids induced by phase separation. The European physical journal E, Soft matter. 2008;25(4):403-13.
43. Alley SH, Ces O, Barahona M, Templer RH. X-ray diffraction measurement of the monolayer spontaneous curvature of dioleoylphosphatidylglycerol. Chemistry and physics of lipids. 2008;154(1):64-7.
44. Illingworth J, Kittler J. A survey of the Hough transform. Computer vision, graphics, and image processing. 1988;44(1):87-116.
45. Jiao J, Rhodes DG, Burgess DJ. Multiple Emulsion Stability: Pressure Balance and Interfacial Film Strength. Journal of Colloid and Interface Science. 2002;250(2):444-50.
46. Blecua P, Lipowsky R, Kierfeld J. Line tension effects for liquid droplets on circular surface domains. Langmuir. 2006;22(26):11041-59.
47. Venkatesan GA, Taylor GJ, Basham CM, Brady NG, Collier CP, Sarles SA. Evaporation-induced monolayer compression improves droplet interface bilayer formation using unsaturated lipids. Biomicrofluidics. 2018;12(2):024101.

48. Ren H, Xu S, Wu S-T. Effects of gravity on the shape of liquid droplets. Optics Communications. 2010;283(17):3255-8.
49. Bolognesi G, Friddin MS, Salehi-Reyhani A, Barlow NE, Brooks NJ, Ces O, et al. Sculpting and fusing biomimetic vesicle networks using optical tweezers. Nature communications. 2018;9(1):1882.
50. Petelska AD, Figaszewski ZA. Interfacial tension of the two-component bilayer lipid membrane modelling of cell membrane. Bioelectrochemistry and Bioenergetics. 1998;46(2):199-204.
51. Petelska AD. Interfacial tension of bilayer lipid membranes. Central European Journal of Chemistry. 2011;10(1):16-26.
52. Elani Y, Gee A, Law RV, Ces O. Engineering multi-compartment vesicle networks. Chem Sci. 2013;4(8):3332-8.
53. Fujita M, Onami S. Cell-to-cell heterogeneity in cortical tension specifies curvature of contact surfaces in Caenorhabditis elegans embryos. PloS one. 2012;7(1):e30224.
54. Contreras FX, Sánchez-Magraner L, Alonso A, Goñi FM. Transbilayer (flip-flop) lipid motion and lipid scrambling in membranes. FEBS letters. 2010;584(9):1779-86.
55. Liu J, Conboy JC. 1,2-diacyl-phosphatidylcholine flip-flop measured directly by sum-frequency vibrational spectroscopy. Biophysical Journal. 2005;89(4):2522-32.
56. Nakano M, Fukuda M, Kudo T, Matsuzaki N, Azuma T, Sekine K, et al. Flip-Flop of Phospholipids in Vesicles: Kinetic Analysis with Time-Resolved Small-Angle Neutron Scattering. The Journal of Physical Chemistry B. 2009;113(19):6745-8.
57. Arai N, Akimoto T, Yamamoto E, Yasui M, Yasuoka K. Poisson property of the occurrence of flip-flops in a model membrane. The Journal of Chemical Physics. 2014;140(6):064901.
58. Vaz WLC, Goodsaid-Zalduondo F, Jacobson K. Lateral diffusion of lipids and proteins in bilayer membranes. FEBS letters. 1984;174(2):199-207.
59. Needham D, Zhelev DV. Lysolipid exchange with lipid vesicle membranes. Annals of biomedical engineering. 1995;23(3):287-98.
60. Guiselin B, Law JO, Chakrabarti B, Kusumaatmaja H. Dynamic Morphologies and Stability of Droplet Interface Bilayers. Physical review letters. 2018;120(23):238001.